\documentclass[conference]{IEEEtran}
\IEEEoverridecommandlockouts

\usepackage{tikz}
\usepackage{textcomp}
\usepackage{hyperref}
\usepackage{lipsum}

\ifCLASSINFOpdf
\else
\fi
\usepackage{algorithm}
\usepackage{algorithmic}
\usepackage{url}

\usepackage[utf8]{inputenc}

\begin{document}

%
\title{A Microservices Architecture for Distributed Complex Event Processing in Smart Cities} 

\author{\IEEEauthorblockN{Fernando Freire Scattone, Kelly Rosa Braghetto}
\IEEEauthorblockA{Department of Computer Science, Institute of Mathematics and Statistics, University of S\~ao Paulo\\
Rua do Mat\~ao, 1010 -- 05508-090 -- S\~ao Paulo -- Brazil\\
Email: \{ffs,kellyrb\}@ime.usp.br}
}


%


\maketitle


\begin{abstract}
A considerable volume of data is collected from sensors today and needs to be processed in real time. Complex Event Processing (CEP) is one of the most important techniques developed for this purpose. In CEP, each new sensor measurement is considered an event and new event types can be defined based on other events occurrence. There exists several open-source CEP implementations currently available, but all of them use orchestration to distribute event processing. This kind of architectural organization may harm system resilience, since it relies on a central core (i.e. the orchestrator). Any failures in the core might impact the whole system. Moreover, the core can become a bottleneck on system performance. In this work, a choreography-based microservices architecture is proposed for distributed CEP, in order to benefit from the low coupling and greater horizontal scalability this kind of architecture provides. 
\end{abstract}


%
\IEEEpeerreviewmaketitle

\section{Introduction}
\label{sec:introduction}


Over the last few years, the concept of Smart Cities has become increasingly popular. A city can be considered smart if investments made in human and social capital and in its transport and communication infrastructure result in a sustainable economic growth and improve citizens quality of life, with a good management of natural resources by participative governance \cite{doi:10.1080/10630732.2011.601117}. Investments made in physical infrastructure usually translate to an implementation of Information and Communication Technology, such as sensors and actuators, on different sectors of the city~\cite{Kitchin2014}. Sensing devices collect a large quantity of valuable data that can be used to improve city services. 

In several applications, real-time data analysis is more valuable than historical data analysis. Detecting fire in a building or detecting a car crash that just happened are some situations that can be better handled the faster they are identified. Tools able to process data collected from a smart city are required to detect these situations in real time, and they should have minimal latency and maximum throughput to achieve this purpose.

Complex Event Processing (CEP) is a powerful resource to process this kind of data. CEP considers each new data input as an event, and allows the use of a number of operators to process events and generate new events when defined patterns are found \cite{Etzion:2010:EPA:1894960}.
However, currently available open-source CEP tools use an orchestration-based architecture to distribute event processing. This kind of architecture concentrates all main operations on a small core part of the application. This impairs system resilience, since a failure in the core part can lead to the system's breakdown. 

Real-time services for smart cities should prioritize availability, because different city applications can depend on these services to work. Therefore, a smart-city service with real-time requirements should be able to deal with local failures without affecting the rest of the system. Moreover, it should also prioritize scalability, since the service usage tends to increase as the city inhabitants increase in number. Studies show city density is expected to grow in the foreseeable future\footnote{\url{www.oecd-ilibrary.org/content/publication/527e5125-en/}}.

In this work, we propose a choreography-based microservices architecture for distributed CEP, in order to benefit from the low coupling and greater horizontal scalability this kind of organization provides. In a choreography, involved parts inherently know what to do and how to react to events received, which lowers the system coupling while increasing its resilience and, in turn, its availability.

A prototype system which implements the proposed architecture in under development. A study case based on a car sharing service will be used to evaluate the system performance in terms of latency and throughput in large scale experiments. The system will also be used to support the study of different load balancing algorithms for distributed CEP. This system will be integrated with the free-software, smart-city platform InterSCity \cite{smartgreens17}. 

\section{Background}
\label{sec:Concetps}

\subsection{Complex Event Processing}
Complex Event Processing (CEP) is a technique to detect specific situations in real time from occurrence of certain input data patterns. The input can be sensor measurements, messages or any other data. 
An \textit{event} can be defined as any piece of data that enters the system or is detected by it. Events are grouped in \textit{event types}.
Events belonging to the same \textit{event type} have the same attributes and are generated by the same source.

A CEP system is composed of three main entities: \textit{Event Producers}, responsible for sending events to the system, \textit{Event Consumers}, responsible for consuming events from the system, and \textit{Event Processing Agents}, or EPAs, responsible for detecting the occurrence of events of given event types. 
Eight main operators can be used to define event types in EPAs: \textit{Filtering}, \textit{Projection}, \textit{Translation}, \textit{Division}, \textit{Enrichment}, \textit{Aggregation}, \textit{Composition}, and \textit{Pattern Detection} \cite{Etzion:2010:EPA:1894960}. Some of these operators are stateless and generate an event for every input event they consume. However, other operators need to wait a number of incoming input events to generate new ones, such as Aggregation, which needs to collect events of the same type. An event type may require the use of more than one operator in order to be detected. 

An example illustrating the use of several operators is a peer-to-peer ridesharing application. Through the application, a client sends an event to the CEP system to call a ride. Then, the CEP system includes client information and preferences on the event and sends it to available drivers. Every driver who chooses to respond positively to the call responds with another event, so clients can choose from several drivers. The system can then generate distinct events to signalize a late arrival or a successful delivery. After the ride, both driver and client can grade each other. The system accumulates grades over time to calculate average grades. In this scenario, the filter operator is used to select available drivers, while the aggregation operator is used to calculate the clients and drivers final grades.


The EPAs input are data streams, i.e. open-ended sets of events. For this reason, stateful operators need to separate incoming events according to their \textit{context}, so only events with specific characteristics are considered in the event detection. Different kinds of context can be used, such as \textit{temporal} (to group events in time windows, according their occurrence time), \textit{spatial} (to partition events by their location), or semantic (to put in a same partition events that relate to a particular value for a given attribute).  

To balance and scale out event processing in a distributed CEP system, stateful operators may need to be replicated in different nodes or moved from a node to another. This can only be done by replicating the operators state, which depends on their defined context. Thus, event types whose operators must use context are the most challenging ones to relocate.



Currently, three main open-source CEP tools have community support: Esper, Drools Fusion and Siddhi\footnote{\url{www.espertech.com/}, \url{www.drools.org/}, \url{github.com/wso2/siddhi/}}. They are offered as libraries which have methods for defining event processing and subsequent actions upon detection in different specific languages. In order to implement them in a distributed manner, the most common approach is to use them coupled with the stream processing tool Apache Storm\footnote{storm.apache.org/}. In Storm, event processing distribution is defined as a directed acyclic graph, where each node represents an environment for event processing and each edge an event flow. However, Storm does not allow node  creation or destruction at runtime, limiting the workload balance possibilities for relocating event types.


\subsection{Microservices and scalability}
In order to balance event processing workload dynamically, the approach proposed in this paper uses virtualized \textit{microservices} as event processing nodes. Microservices can be defined as \textit{small}, \textit{autonomous} services that can work well together:  \textit{small} because their operations must be contained in a specific domain of operations, and \textit{autonomous} because they should be able to be executed independently of their system~\cite{Newman:2015:BM:2904388}.





A microservices architecture can be organized in two ways: using \textit{orchestration}, where a central microservice is responsible for coordinating all main system operations, or using \textit{choreography}, where each microservice is responsible for its own operations and needs to communicate with others to perform complex operations and meet a common purpose. The use of choreography improves the system availability, since in a choreography-based architecture there is no single point of failure that impacts the whole system.




Techniques for achieving scalability in stateful microservices are the same ones used in distributed database management systems (DBMSs): \textit{replication} and \textit{sharding}.
With replication, the whole system state is replicated in several nodes. This leads to great availability, but it comes at the cost of possible stale data accesses while new entries are copied to the replicas. With sharding, the state is partitioned and each partition is stored in a different location (node). Sharding requires the use of a balancing algorithm for determining the location of new entries. This ensures consistency but lowers availability, since each entry has only one copy and a single access point. However, the two techniques can be used combined to provide better robustness.


\section{Related Works}

Several works have already addressed scalability in CEP by proposing workload balancing algorithms for CEP operators allocation in distributed environments.

Isoyama et al.~\cite{Isoyama:2012:SCE:2335484.2335498} approach places together event types processing that have the same input event types. It also uses shared state among different EPAs to distribute stateful operators processing. 

The mechanism proposed by Pathak and Vaidehi~\cite{7129184} checks for every new event type definition if it is not already being detected and sets a maximum number of distinct event types for each CEP instance to process. Each new event type definition is stored together with other metadata (e.g. its own input event types and the event types to which the new type is an input). This metadata is used to select the instance that will process the new event type.


Martins et al. \cite{6906776} introduced the Audy architecture, which distributes event processing based on resource utilization of the CEP instances. Input and output event queues are used to monitor the resource usage. They proposed the detection of two different situations in an instance: light overload, which triggers the request for relocating the detection of the last event type registered, and hard overload, which stops the detection of the last event type  registered.
This technique is adopted in order to increase the system availability, even if the processing of some event types is discarded.

Some of the above-mentioned works \cite{Isoyama:2012:SCE:2335484.2335498,7129184} have focused in scalability, while others \cite{6906776} have also considered the system availability. As discussed in Section~\ref{sec:introduction}, a CEP system for smart cities  should consider both requirements as equally important.

 

\section{A Microservices Architecture to Support Distributed CEP}

This paper proposes a new distributed CEP architecture based on two microservices: \textit{CEP-Worker} and \textit{CEP-Cataloger}. The CEP-Worker microservice is responsible for event processing, load balancing, and scaling the system out/in (by creating/destroying CEP-Worker instances).
The CEP-Cataloger microservice provides an interface to  create, update and delete the event types to be detected by the CEP system.  This architecture was designed to be implemented as a service in a smart city plataform (such as the InterSCity platform \cite{smartgreens17}), so that users can define event types of interest in a city and set an address as web-hook to be called when they are detected. 



\subsection{Choreography-based CEP distribution}

Each CEP-Worker instance is responsible for detecting event types and send detected events through asynchronous brokers to other instances that may use it as input events. Every time a new event type is registered to be detected in the CEP system, it is allocated to the instance with less resource usage. 
Each CEP-Worker instance must constantly monitor its workload and resource consumption. 

When its incoming events flow is too high, a CEP-Worker instance searches for another instance whose resource consumption is below the limit, to relocate some of its own event types in this instance. If all other instances are fully loaded, a new instance of the CEP-worker is created  (i.e. the system scales out) to receive the relocated event types from the overloaded instance.
This way, CEP-Worker instances communicate among themselves and collaborate in a choreographic approach in order to balance their workload distribution. 

Two different strategies may be used by an overloaded CEP-Worker instance to choose the event types to be relocated
 in other instance. The first strategy (Algorithm \ref{alg:relocation_similarity}) considers the similarity among the input types of the event types in the overloaded instance. Event types whose input event types differ most from the others will be selected for relocation. When the input types of event types in the overloaded instance are all the same, event types to be relocated can be chosen randomly or according their resource utilization (Algorithm \ref{alg:relocation_resource_usage}), measured in function of the context definition they use. In this second strategy, event types whose contexts use more state are relocated last, while stateless event types are relocated first since they can be transferred to other instances more easily. 

If the input events flow for a CEP-Worker instance is too low, it searches for another instance with low resource usage to relocate its own event types. After the relocation is concluded, the instance self-destructs (i.e. the system scales in). 

During the relocation of an event type from a CEP-Worker instance to another, the two instances must keep detecting the same event type until it is acknowledged that their detections generate similar results. This way, the system can prevent event loss in case of failures in relocation.

\begin{algorithm}[tb]
\begin{algorithmic}[1]
\caption{SearchTypesByInputSimilarity(instance)}
\label{alg:relocation_similarity}
\STATE \COMMENT{\textit{MAX\_FLOW = maximum event flow threshold}}
\STATE F $\leftarrow$  FlowRateOfIncomingEvents(instance)
\STATE usesByInput $\leftarrow$ \{\} \COMMENT{\textit{this is a dictionary}}  
\STATE eventTypesToRelocate $\leftarrow$ []  \COMMENT{\textit{this is a list}} 
\FOR{ each inputEType $\in$ instance.inputEventTypes} 
\STATE {usesByInput[inputEType] $\leftarrow$ 0 }
\FOR{ each eType $\in$ instance.eventTypes}
\IF { inputEType $\in$ eType.inputs}
\STATE {usesByInput[inputEType]++}
\ENDIF
\ENDFOR
\ENDFOR
\STATE listUsesByInput $\leftarrow$ GetListOfPairs(usesByInput) \COMMENT{\textit{generates a list of pairs (input event type, number of uses)}}
\STATE orderedUsesByInput $\leftarrow$ OrderByUses(listUsesByInput)
\WHILE{F $>$ MAX\_FLOW} 
\STATE lessUsedInput $\leftarrow$ RemoveFirst(orderedUsesByInput) 
\IF {lessUsedInput[2] $<$ sizeof(instance.eventTypes)}
\FOR{ each eType $\in$ instance.eventTypes}
\IF {  lessUsedInput[1] $\in$ eType.inputs }
\STATE eventTypesToRelocate $\leftarrow$ eventTypesToRelocate + [eType]
\STATE F = F - eType.flow
\ENDIF
\ENDFOR

\ELSE 
\STATE randomEType $\leftarrow$ SelectRandomly(instance.eventTypes $-$ eventTypesToRelocate)
\STATE eventTypesToRelocate $\leftarrow$ eventTypesToRelocate + [randomEType]
\STATE F = F - randomEType.flow
\ENDIF

\ENDWHILE

\RETURN eventTypesToRelocate
\end{algorithmic}
\end{algorithm}


\begin{algorithm}[tb]
\caption{SearchTypesByResourceUsage(instance)}
\label{alg:relocation_resource_usage}
\begin{algorithmic}[1]
\STATE \COMMENT{MAX = maximum resource usage threshold}
\STATE IC $\leftarrow$  InstanceConsumption(instance)
\STATE ECs $\leftarrow$  [] \COMMENT{\textit{list of pairs (event type, consumption)}}
\STATE eventTypesToRelocate = []
\STATE context $\leftarrow$ GetContextDataFromDatabase(instance)
\FOR { each eType $\in$ instance.eventTypes}
\STATE ECs  $\leftarrow$  ECs +  [(eType, EventTypeConsumption(context[eType])] 
\ENDFOR
\STATE orderedECs $\leftarrow$ OrderByConsumptionDesc(ECs)

\WHILE{IC $>$ MAX} 
\STATE biggestEC $\leftarrow$ RemoveFirst(orderedECs) 
\STATE eventTypesToRelocate $\leftarrow$ eventTypesToRelocate + [biggestEC[1]] 
\STATE IC $\leftarrow$ IC - biggestEC[2]
\ENDWHILE
\RETURN eventTypesToRelocate
\end{algorithmic}
\end{algorithm}

The CEP-Cataloger microservice is responsible for registering all event types in the system. It provides an API for developers to register new event types and web-hooks (web pages which are called by CEP-Cataloger when specified events are detected). CEP-Cataloger uses a DBMS to store all event type metadata to be used when event types are relocated in CEP-Worker instances. The DBMS must be scalable by replication in order to ensure availability, since CEP-Worker instances will query event types metadata frequently. 
Since the event types metadata is stored in the DBMS, in case of an instance failure the affected event types can be relocated to another instance as soon as the failure is detected.

\subsection{Proposed implementation for the architecture}

The CEP system described in this work is been developed as part of the activities of INCT of the Future Internet for Smart Cities (InterSCity)\footnote{interscity.org/}. In accordance to the InterSCity philosophy and to facilitate the integration with its smart city platform \cite{smartgreens17}, open-source software tools were chosen to implement the system: Esper as CEP engine, Ruby on Rails for the stateless microservice CEP-Cataloger, RabbitMQ for asynchronous event transmission, Docker for microservice environment virtualization, and Redis\footnote{\url{rubyonrails.org/}, \url{www.rabbitmq.com/}, \url{docs.docker.com/}, \url{redis.io/}} as scalable DBMS. Esper was chosen because it is the most popular CEP engine and is used in most part of the related works. The other tools were chosen based on their performance and usage on the InterSCity smart city platform.


The implementation of the proposed architecture is an ongoing work. CEP-Worker basic functionalities, new event types registry in execution time, and asynchronous event transmission have already been implemented. Next steps include the implementation of CEP-Cataloger and CEP-Worker's event type relocation algorithms.

\subsection{System evaluation}

In order to evaluate the proposed architecture, a study case based on a peer-to-peer ridesharing application will be implemented using CEP operators, similarly to the one presented on Section \ref{sec:Concetps}. Drivers will have different payment options, considering how the ride is charged (e.g. by time elapsed, by distance, and by driver offer). Additionally, the system will store client preferences (minimum driver grade, car age, shared ride usage, etc.) in order to only offer vehicles that match clients criteria. The study case will use both stateful and stateless operators in order to evaluate the performance of load balancing approaches.

An experiment will be executed simulating car traffic in a large city during six continuous hours, capturing an increase and decrease in traffic flow to test the system elasticity. Vehicles position data will be generated by the InterSCSimulator \cite{mabs2017}, a large scale mesoscopic simulator based on actors. All other data, including client and driver data, and events, including ride calling events and driver responding events, will be generated by the system itself.


Three main measurements will be taken during  experiments: latency on event types detection, events throughput, and number of instances in execution. From the measures, the workload distribution approaches will be evaluated to check if event processing should be distributed based on event types similarity of input types, on resource usage, or on a combination of both. 

\section{Conclusion}

CEP is a powerful technique to analyze real-time data, but most of current open-source CEP tools do not offer support to a large scale dynamic distribution of workload. 
By using microservices to distribute CEP, where instances communicate with each other through an asynchronous broker, the whole system becomes less coupled. Differently from the approach of Martins et al. \cite{6906776} which always relocate the last event type assigned to an instance, this work proposes two search algorithms to identify the event types that should be relocated and chooses instances with least resource usage to place them.

As future work, the evaluation of the two search algorithms will be performed in order to provide a scalable CEP system able to be used by a large number of citizens, through various smart city applications running over a same elastic computational infrastructure. 
A study case based on car sharing services will be used to evaluate the proposed system and architecture. A simulator will generate car traces for a large city, which will be used to test system latency,  throughput, and resource consumption.


\section*{Acknowledgment}
This research is part of the INCT of the Future Internet for Smart Cities funded by CNPq, proc. 465446/2014-0, CAPES proc. 88887.136422/2017-00, and FAPESP, proc. 2014/50937-1. Fernando Freire Scattone is supported by CAPES.




%

\bibliographystyle{IEEEtran}
\bibliography{bare_conf.bib}

\end{document}